\documentclass[journal=langd5,manuscript=article]{achemso}
\usepackage[version=3]{mhchem} 
\usepackage{natbib}
\usepackage{url}
\usepackage{appendix}
\usepackage{subcaption}

\usepackage{epstopdf}
\usepackage{mathtools}
\usepackage{textgreek}

\usepackage{multirow}
\usepackage{multicol}
\usepackage{graphicx}

\usepackage{caption}

\usepackage{amssymb}
\usepackage{float}
\usepackage[export]{adjustbox}
\usepackage{slashbox}

\usepackage{xcolor}

\author{Hari Govindha A.}
\affiliation{Department of Mechanical and Aerospace Engineering, Indian Institute of Technology Hyderabad, Kandi - 502284, Telangana, India}

\author{Saravanan Balusamy}
\affiliation{Department of Mechanical and Aerospace Engineering, Indian Institute of Technology Hyderabad, Kandi - 502284, Telangana, India}
\email{saravananb@mae.iith.ac.in}
\author{Sayak Banerjee}
\affiliation{Department of Mechanical and Aerospace Engineering, Indian Institute of Technology Hyderabad, Kandi - 502284, Telangana, India}
\email{sayakb@mae.iith.ac.in}
\author{Kirti Chandra Sahu}
\affiliation{Department of Chemical Engineering, Indian Institute of Technology Hyderabad, Kandi - 502284, Telangana, India}
\email{ksahu@che.iith.ac.in}

\title{Intricate Evaporation Dynamics in Different Multi-Droplet Configurations}

\begin{document}

\maketitle
\begin{abstract}
We experimentally investigate the evaporation dynamics of an array of sessile droplets arranged in different configurations. Utilizing a customized goniometer, we capture side and top view profiles to monitor the evolution of height, spread, contact angle, and volume of the droplets. Our results reveal that the lifetime of a droplet array surpasses that of an isolated droplet, attributed to the shielding effect induced by neighbouring droplets, which elevates the local vapour concentration, thereby reducing the evaporation rate. We found that lifetime increases as droplet separation distance decreases at a fixed configuration and substrate temperature. It is observed that the lifetimes increase with the number of droplets. We observe a decrease in lifetimes, following a power law trend with increasing substrate temperature, with the shielding effect diminishing at higher substrate temperatures due to natural convective effects. We also observe a generalised behavior for the centrally placed droplet across various separation distances and substrate temperatures. This arises from different droplet configurations and substrate temperatures, which modify the local vapour concentration around the droplets without significantly impacting the contact line dynamics. Additionally, the experimental results are compared with a diffusion-based theoretical model that incorporates the evaporative cooling effect to predict the lifetime of the central droplet within the array. We observe that the theoretical model satisfactorily predicts the lifetime of the droplet at room temperature. However, for high-temperature cases, the model slightly overpredicts the evaporative lifetimes.
\end{abstract} 


\section{Introduction}
\label{sec:intro}

The evaporation of surface-adhering droplets is widely encountered across various fields, including inkjet printing \cite{de2004inkjet,tekin2004ink,soltman2008inkjet,park2006control}, DNA microarray fabrication \cite{dugas2005droplet,lee2006electrohydrodynamic}, disease transmission \cite{balusamy2021lifetime,bhardwaj2020likelihood}, coating technologies \cite{kim2016controlled,pahlavan2021evaporation}, spray and hotspot cooling \cite{kim2007spray,ruan2019effects,cheng2010active}, and microfluidics \cite{deegan1997capillary}, to name a few. While earlier research predominantly concentrated on the evaporation of individual sessile droplets, practical scenarios often involve droplets arranged in arrays or dispersed randomly on a surface. In such situations, understanding the collective impact of neighbouring droplets on their evaporation dynamics becomes crucial. The presence of other droplets in the proximity of a droplet is characterized by a decrease in the evaporation rate due to the increase in local humidity and the shielding effect \cite{schofield2020shielding}. Below, we present a brief literature review summarizing key findings on the evaporation of sessile drops on surfaces maintained at different temperatures.

The pioneering investigations of \citet{birdi1989study,picknett1977evaporation} on the evaporation of sessile water droplets on hydrophilic substrates at room temperature reveal a constant evaporation rate. \citet{picknett1977evaporation} identified two primary modes in the evaporation process, namely the constant contact radius (CCR) and constant contact angle (CCA) modes. It was observed that the evaporation primarily occurred in the CCR mode until the droplet reached the receding contact angle, followed by a transition to the CCA mode. They also developed a theoretical model to predict the evaporation rate of the droplet. The distribution of vapour concentration, the evaporative flux above the droplet surface, and the effect of Marangoni stresses were investigated with the help of a model in the framework of the Finite Element Method (FEM) and a series of experiments \cite{hu2002evaporation,hu2005analysis}. \citet{shin2009evaporating} experimentally investigated the evaporation dynamics of water droplets on surfaces with varying hydrophilicities, revealing distinct contact line dynamics for hydrophilic, hydrophobic, and superhydrophobic surfaces. \citet{yu2012experimental} conducted experiments on the evaporation of water droplets on Teflon and Polydimethylsiloxane (PDMS) surfaces and found the presence of CCR, CCA, and mixed modes in the evaporation process. 
All of these studies focus on the evaporation of droplets at ambient temperature.

An elevated substrate temperature introduces distinct evaporation phenomena in droplets compared to those observed at room temperature due to the presence of thermocapillary flow. Using infrared (IR) thermography, Girard et al. \cite{girard2010infrared, girard2008effect} investigated the evaporation rate of water droplets and examined the temperature variation between the apex and contact line of the droplet. Their findings suggested that the Marangoni flow had a negligible effect on the evaporation rate of the droplets. \cite{gurrala2019evaporation} conducted both experimental and theoretical investigations to examine the influence of composition and substrate temperatures on the evaporation of pure and binary sessile droplets. They developed a theoretical model that incorporates diffusion, free convection, and passive transport due to the free convection of air. At elevated substrate temperatures, they observed a non-monotonic trend in the lifetime of the droplet with increasing ethanol concentration in the binary mixture. Interestingly, they found that the evaporation dynamics for different compositions at high substrate temperatures exhibited a self-similar trend, demonstrating a constant normalized volumetric evaporation rate throughout the evaporation process. \citet{sobac2012thermal} conducted experiments to explore the evaporation of ethanol droplets at different temperatures on hydrophilic and hydrophobic surfaces and developed an evaporation model based on quasi-steady, pure diffusion-driven principles. \citet{brutin2011infrared} used an infrared (IR) camera to visualize the thermal patterns during the evaporation of sessile droplets of different semi-transparent liquids and observed that the surface instability depends on the fluid considered. Utilizing IR thermography, \citet{sefiane2013thermal} demonstrated that hydrothermal waves span the entire droplet volume, with thermal patterns influencing heat flux distribution throughout the droplet. \citet{zhong2017stable} found that elevating the substrate temperature amplifies thermocapillary instabilities and temperature variations within hydrothermal waves. \citet{dash2014droplet} reported the evaporation of sessile water droplets on hydrophobic and superhydrophobic surfaces at higher substrate temperatures ($40^\circ$C - $60^\circ$C). 
Upon comparing experimental results with theoretical predictions, they found that the vapour-based diffusion model only agrees with the evaporation rates on hydrophobic surfaces. In addition to these studies, several researchers have explored the impact of various factors on the evaporation of sessile droplets, including the thermal conductivity of the substrate \cite{dunn2009strong,sobac2012thermal}, thermal and solutal Marangoni convection effects \cite{van2021marangoni,kim2016controlled}, changes in contact line dynamics with particle suspensions \cite{orejon2011stick,deegan2000pattern}, substrate roughness, geometries, and inclinations \cite{pittoni2014uniqueness,gurrala2022evaporation,katre2022experimental}. These investigations employ a combination of analytical, numerical, and experimental techniques.

A few researchers have also investigated the simultaneous evaporation phenomenon of multiple droplets. \citet{carrier2016evaporation} conducted experimental and theoretical analyses of evaporation rates for single and multiple droplets and observed a decrease in evaporation rates for multiple droplets due to atmospheric saturation from neighbouring drops, a phenomenon similar to the shielding effect observed in dissolution processes. \citet{laghezza2016collective} found a 60\% increase in dissolution time for alcohol droplets immersed in a water pool compared to isolated droplets, which further illustrated the complex evaporation in the case of multiple drops. Similarly, \citet{shaikeea2016insight} analyzed a two-droplet system on a hydrophobic surface and reported a 50\%  reduction in evaporation rates and asymmetrical left and right contact line dynamics during the initial stages of evaporation. \citet{pradhan2015deposition} experimentally investigated internal flow and deposition patterns in two-droplet systems, observing weaker deposition near the droplets due to differential evaporation rates. \citet{khilifi2019study} examined a seven-droplet configuration on a sapphire substrate and found that the lifetime of droplets decreases with increasing inter-droplet distances. This finding was supported by a competitive diffusion-limited asymptotic evaporation model developed by \citet{wray2020competitive}. \citet{hatte2019universal} and \citet{pandey2020cooperative} derived universal evaporation dynamics for three-droplet and $5 \times 5$ two-dimensional (2D) array systems using the scaling of the droplet lifetime across various $L/d$ ratios. Here, $d$ denotes the equatorial diameter of the droplet. \citet{tonini2022analytical} developed analytical models predicting evaporation for two and three-droplet configurations on hydrophilic and hydrophobic surfaces, with more significant evaporation rate reductions observed on hydrophobic substrates than the hydrophilic substrates at a given $L/d$ ratio. \citet{masoud2021evaporation} formulated models predicting diffusive evaporative rates for multiple droplets of different sizes and contact angles. The interferometric techniques \cite{edwards2021interferometric} were used to measure the volumes of various droplets, which is impossible using traditional methods. Using the pattern distortion method, \citet{iqtidar2023drying} tracked the volumes of central and side droplets for different arrangements of multiple droplets and showed similarity in the drying dynamics for all arrangements. \citet{rehman2023effect} and \citet{chen2022predicting} also explored the evaporation dynamics of multiple droplets on hydrophilic and hydrophobic substrates, respectively. They developed theoretical and numerical models to predict the evaporation rates of droplets in different substrate conditions. 

The review discussed above reveals that while significant attention has been devoted to comprehending the evaporation dynamics of individual sessile droplets at both room and elevated substrate temperatures, research on the simultaneous evaporation of multiple droplets has been relatively limited. The existing investigations have predominantly focused solely on room temperature conditions. However, as discussed above, even a single droplet exhibits distinct dynamics at high substrate temperatures. Therefore, the present work concentrates on investigating the evaporation dynamics of multiple droplets at elevated substrate temperatures, which are frequently observed in practical applications \cite{soltman2008inkjet,park2006control,lee2006electrohydrodynamic,balusamy2021lifetime,kim2016controlled,pahlavan2021evaporation,kim2007spray}. Six different droplet configurations, ranging from isolated droplets to arrangements of six droplets, are considered, along with three $L/d$ ratios of $1.2$, $1.6$, and $2.0$ and the substrate temperature is varied from $25^\circ$C to $65^\circ$C. In our experiments, we analyze the evaporation dynamics and estimate the lifetime of droplets in all configurations. However, given constraints in accessing side view profiles of central droplets in cases involving four and six droplets, we focus on presenting the temporal evolution of droplet height, wetting diameter, contact angle, and volume exclusively for configurations with one, three, and five droplets at $L/d$ ratios of $1.6$ and $2.0$. Furthermore, the experimental results are compared with a diffusion-based theoretical model that incorporates the evaporative cooling effect to predict the lifetime of the central droplet within the array.

\section{Experimental Methodology}
\label{sec:expt}
\subsection{Experimental Setup}

\begin{figure}[h]
\centering
\includegraphics[width=0.9\textwidth]{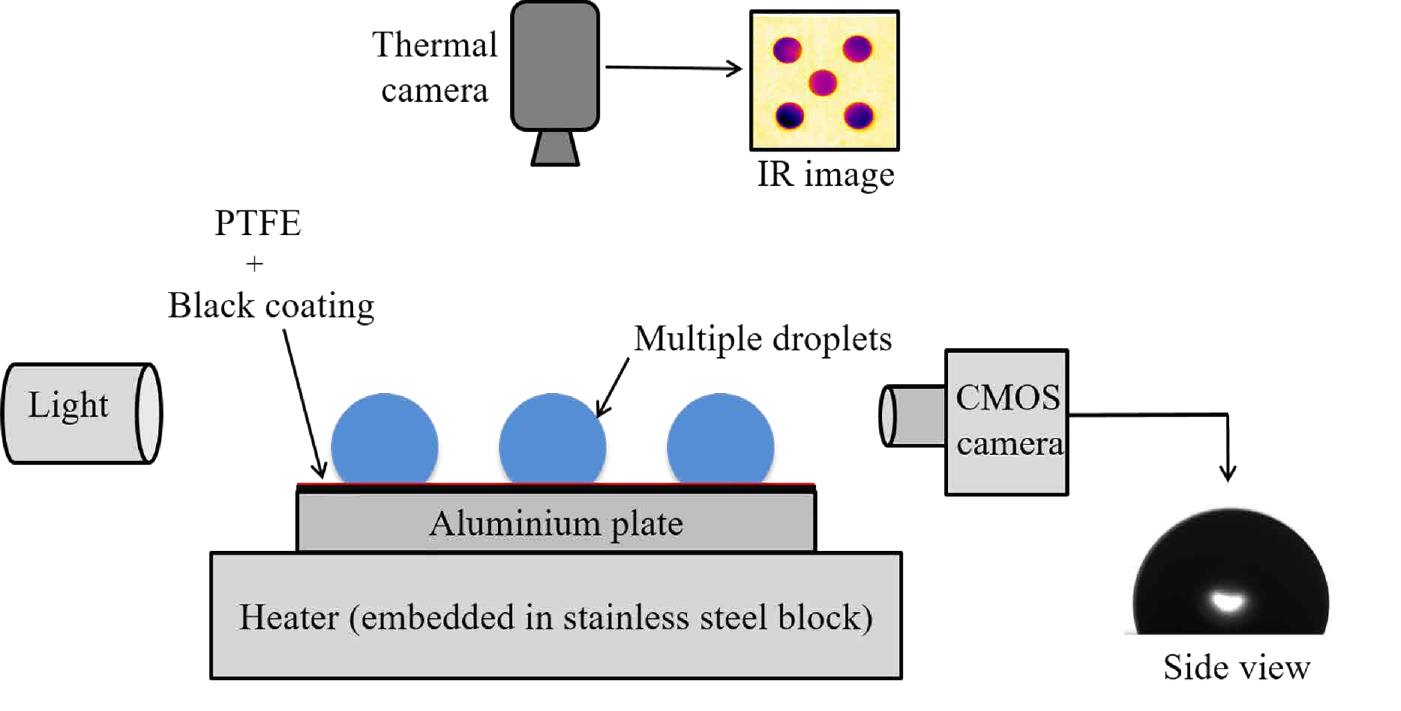}
\caption{Schematic diagram of the experimental setup (customized goniometer).} 
\label{fig:fig1}
\end{figure}

We experimentally investigate the evaporation dynamics of multiple sessile water droplets with various configurations using shadowgraphy and infrared (IR) imaging techniques. The schematic of the experimental setup is shown in Fig. \ref{fig:fig1}. The goniometer unit is customized for our requirements (Make: Holmarc Opto-Mechatronics Pvt. Ltd.). It consists of a multilayered metallic block, a motor-driven pump for dispensing the droplets, a proportional-integral-derivative (PID) controller for regulating the temperature of the substrate, a complementary-metal-oxide-semiconductor (CMOS) camera (Make: Do3Think, Model: DS-CBY501E-H), an infrared camera (Make: FLIR, Model: X6540sc), and an LED light source with a diffuser sheet. The CMOS and IR cameras were used to capture the side and top views of the evaporating droplet, respectively. Note that due to challenges associated with reflection, translating the infrared (IR) data to temperature is difficult. Consequently, in this study, we do not present the temperature distribution on the free surface of the drops. The assembly was placed inside the goniometer box to minimize external environmental disturbances. All experiments were performed at an ambient temperature of $22^\circ$C $\pm$ $1^\circ$C and a relative humidity of $65\pm 5$\%. The relative humidity was measured using a hygrometer (Make: HTC, Model: 288-ATH) fitted inside the goniometer box. The substrate temperature is varied from $25^\circ$C to $65^\circ$C.

The multilayered metallic block consists of (i) a stainless steel base fitted with two electrical heaters operated by the PID controller and (ii) an aluminum plate of size 100 mm $\times$ 80 mm $\times$ 15 mm. The aluminum plate is coated with black paint to minimize the reflections in the IR images. A CMOS camera with a spatial resolution of $1280 \times 960$ pixels recorded the side view of the droplet at ten frames per second (fps), which were used to extract various droplet parameters, such as the wetted diameter ($D$), height ($h$), contact angle ($\theta$) and volume ($V$). The top views of the droplets are recorded with a resolution of $640 \times 512$ pixels at 50 fps in the spectral range of $3$ $\mu$m – $5$ $\mu$m. A polytetrafluoroethylene (PTFE) tape of thickness 100 $\mu$m is pasted on the aluminum plate, which is used as the substrate. The thermal stability and roughness of the PTFE tape were verified for the temperature range considered in this study \cite{katre2022experimental,Govindha2022}. The required substrate temperature was then set with the help of a PID controller, and a K-type thermocouple (Make: OMEGA Engineering Singapore) was used to check whether the substrate attained the steady state temperature before dispensing the droplets. Before performing each experiment, it is ensured that the PTFE substrate is thoroughly cleaned with isopropanol, dried with compressed air, and then pasted onto the aluminum plate. A 50 $\mu$L U-tek (Make: Unitek Scientific Corporation) chromatography syringe (with a piston size of 1.08 mm and fitted with a 21G needle with an inner orifice diameter of 0.514 mm) was connected to the motorized pump to control the volume flow rate, which in turn dispensed droplets of a constant size.

\subsection{Droplet dispensing method}
\begin{figure}[htbp]
\centering
\includegraphics[width=0.9\textwidth]{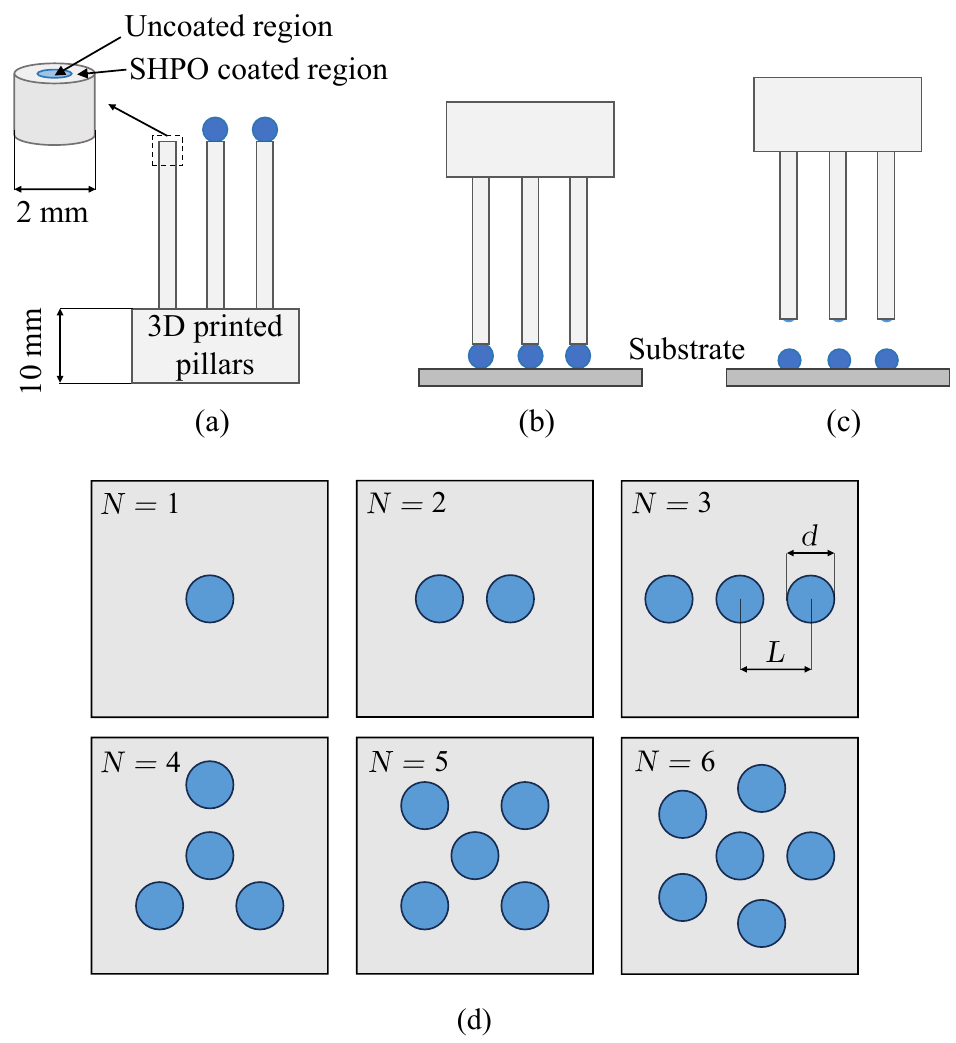}
\caption{Schematic representation of the droplet dispensing method. (a) Placement of individual droplets on pillars. (b) Inverted pillar setup for simultaneous droplet deposition. (c) Deposition of multiple droplets. (d) Various droplet configurations achieved using this process. Here, $N$ signifies the number of droplets arranged in different configurations, $d$ denotes the equatorial diameter of the droplet, and $L$ represents the distance between the central and side droplets.} 
\label{fig:fig2}
\end{figure}

In this study, dealing with multiple droplets at high substrate temperatures presents challenges, particularly regarding non-uniformities in initial volumes among the droplets when individually placed on the heated substrate. To address this, a novel approach is developed, as schematically illustrated in Fig. \ref{fig:fig2}. Initially, a support structure comprising pillars is fabricated using a 3D printer. These pillars are made from a transparent hydrophobic (HPO) polymer resin. Numerous pillars are created based on the desired droplet configurations. Subsequently, the pillars are coated with a superhydrophobic (SHPO) coating (Make: Glaco, Model: Mirror coat zero). Notably, the SHPO coating is removed from the central region of the pillars (uncoated region) to enable fixation of the droplets onto the pillars, as depicted in Fig. \ref{fig:fig2}(a). This setup ensures that the deposited droplets on the pillars come into contact with the HPO polymer resin material, surrounded by a region of SHPO coating. The combination of HPO and SHPO regions prevents roll-off and impedes complete spreading of the droplet on the pillar surface. By restricting the complete spread of the droplet, the contact angle is increased from its equilibrium value, which would otherwise occur on the HPO pillar surface without the SHPO coating. This modification makes the droplet less prone to roll-off, easing its dislodging when brought in contact with the substrate surface, as the entire weight of the droplet is supported by a smaller contact area.

Subsequently, the pillar arrangement is inverted, and the droplets are deposited on the heated substrates simultaneously, as shown in Fig. \ref{fig:fig2}(b). While in the inverted position, the droplets are held in place with the help of the HPO region. Upon contacting the substrate, they begin spreading on it, leaving a slight residue on the pillar, as illustrated in Fig. \ref{fig:fig2}(c). For a given droplet size, the size of the uncoated region should be such that it is smaller than the equilibrium wetting diameter but not so small that the droplet immediately falls when inverting the arrangement. The number of droplets ranges from a single isolated droplet to an arrangement of six droplets, as depicted in Fig. \ref{fig:fig2}(d). The distance between the central and side droplets is denoted by $L$, while $d$ represents the equatorial diameter of the droplet. Different $L/d$ ratios are achieved by varying only the distance between the droplets while keeping $d$ constant at 2.5 mm.

The volumes of the droplets created using this mechanism were approximately (4.8 $\mu$l $\pm$ 0.3 $\mu$l). In our experiments, time $t=0$ denotes when the droplet makes contact with the substrate. The maximum time from placing all the droplets on the pillar to depositing them on the substrate is approximately 100 seconds (for a six-droplet configuration), which is insignificant compared to the evaporation time of the same configuration at room temperature. Following each experiment, the PTFE tape was replaced. We conducted a minimum of three repetitions for each set of experimental conditions.

\subsection{Post-processing}

To extract side-view profiles of the droplets, we conducted post-processing on the side-view images captured by the CMOS camera using a custom program developed within the \textsc{Matlab}$^{\circledR}$ framework. This process involved employing a median filtering technique to remove random noise and an unsharp masking technique to enhance image sharpness and gradients. Subsequently, the filtered image was converted into a binary format using an appropriate threshold to distinguish between the background and the droplet boundary. Any holes within the droplet boundary were filled, and the droplet reflection was eliminated. We utilized a \textsc{Matlab}$^{\circledR}$ function to trace the droplet contour, from which droplet parameters were measured. This post-processing procedure closely resembles what was described by \citet{gurrala2019evaporation}.
For infrared images, intensity data were converted into a temperature field. Boundary extraction was performed using a Convolutional Neural Network based on the U-net architecture, as detailed in Ref. \citet{katre2021evaporation}. The U-net design incorporated data augmentation by elastically deforming annotated input photos, enhancing the network's utilization of available annotated images. The network was trained using 40 manually annotated grayscale infrared images, with training conducted on a computer equipped with a GPU (NVIDIA Quadro P1000). Subsequently, the trained network was employed to extract binary masks and droplet boundaries from the infrared images. Finally, a \textsc{Matlab}$^{\circledR}$ code was utilized to remove background noise, and the top-view profiles of the evaporating droplets at various time intervals were analyzed.

\section{Results and Discussion}\label{sec:results}

\subsection{Multiple droplet lifetime ($t_e$)}

We investigate the evaporation dynamics of an array of sessile water droplets on a hydrophobic substrate arranged in different configurations. In all the configurations considered (except for single and two-droplet configurations), we have a centrally placed droplet surrounded by side droplets, which are kept at a definite distance from the central droplet. In this study, we considered three different $L/d$ ratios of $1.2$, $1.6$, and $2.0$. The $L/d$ ratio indicates the proximity of the droplets relative to their equatorial diameter. Apart from varying $L/d$ ratios, the experiments are performed at three different substrate temperatures of $T_s = 25^\circ$C, $45^\circ$C, and $65^\circ$C. Table \ref{table:T1} presents the lifetime $(t_e)$ of the central droplet in various configurations, $L/d$ ratios, and at different substrate temperatures $(T_s)$.

\begin{table}
\centering
\caption{The lifetime $(t_e)$ of the central droplet in different configurations (in second) for different values of $L/d$ and substrate temperature $(T_s)$. The maximum deviation is $\pm~7\%$ for the 3-droplet configuration with $L/d = 1.6$ at $T_s = 25^\circ$C.}
\label{table:T1}
\hspace{2 mm}\\
\begin{tabular}{|c|c|c|c|c|c|c|c|c|c|}\hline
Substrate temperature &\multicolumn{3}{|c|}{$25^\circ$C} & \multicolumn{3}{|c|}{$45^\circ$C} &\multicolumn{3}{c|}{$65^\circ$C}\\
\hline
\multirow{2}{*}{\backslashbox{$N$}{$L/d$}}& $1.2$ &  $1.6$ & $2.0$ & $1.2$ & $1.6$ &$2.0$ &  $1.2$ & $1.6$ &$2.0$\\ 
	
  &  &  & & &  & &  &  & \\ 
 \hline
1 &  \multicolumn{3}{|c|}{3720}  & \multicolumn{3}{|c|}{712}  & \multicolumn{3}{|c|}{320}\\


\hline

2 & 4383 & 4160  & 3975 & 836  & 794 & 746& 357  & 342 & 336  \\
 \hline

3 & 5074  & 4629  & 4354  & 908  & 851 & 795 &  385 & 366 & 352 \\
 \hline

4 & 5646  & 5214  & 4755  & 984  & 914  & 858& 416 & 386  & 365 \\
 \hline

5 & 6394 & 5620  & 5275  & 1090  & 988  & 909 & 439 &404  & 378 \\
 \hline

6 & 7104  & 6465  & 5822  & 1154  & 1052  & 974& 474  & 423  & 390  \\
 \hline
\end{tabular}
\end{table}

\begin{figure}[h]
\centering
\hspace{0.5cm}\\
\hspace{-0.2cm}\includegraphics[width=0.45\textwidth]{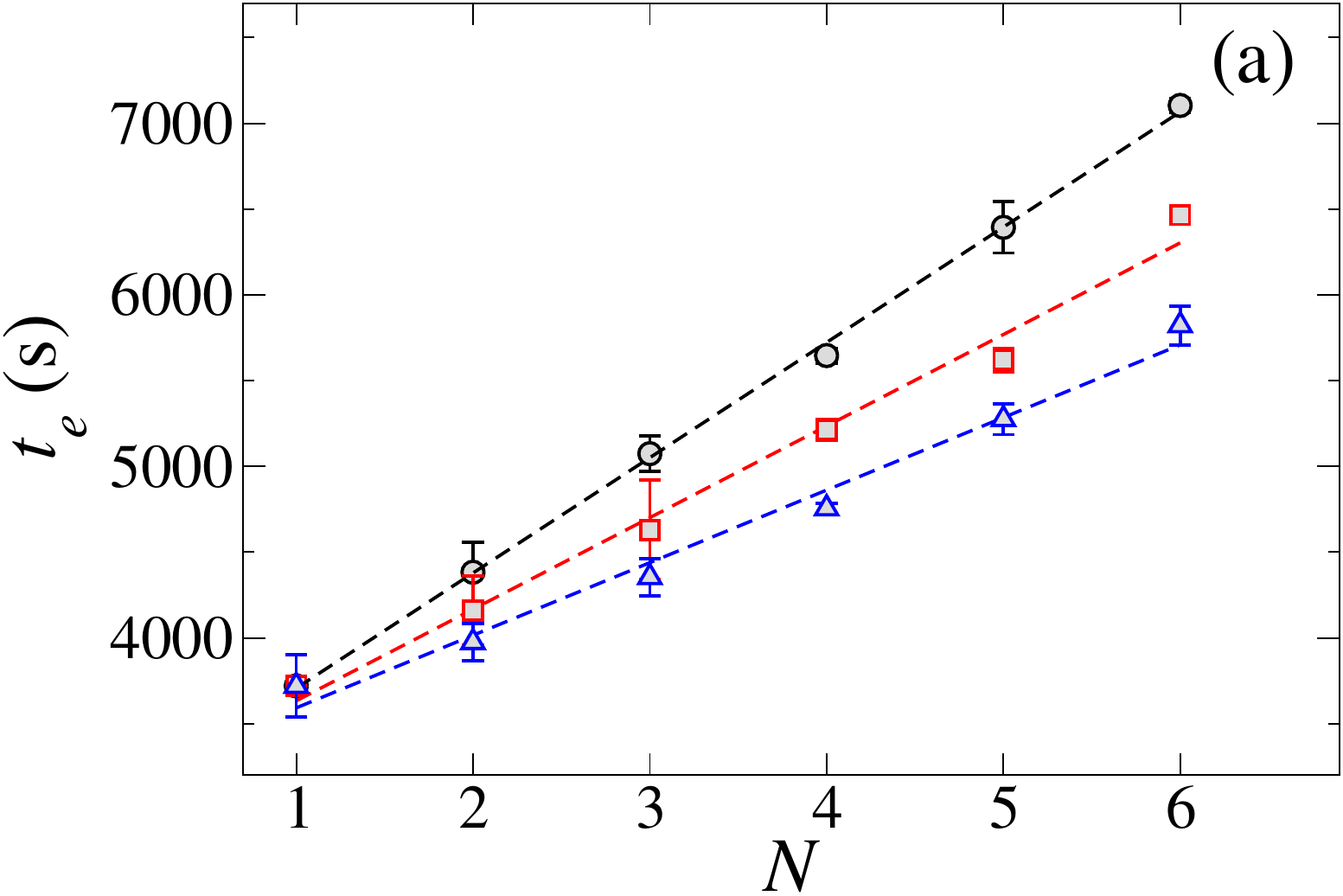} \hspace{2mm}\\
\hspace{0.5cm} \hspace{7.1cm} \\
 \includegraphics[width=0.45\textwidth]{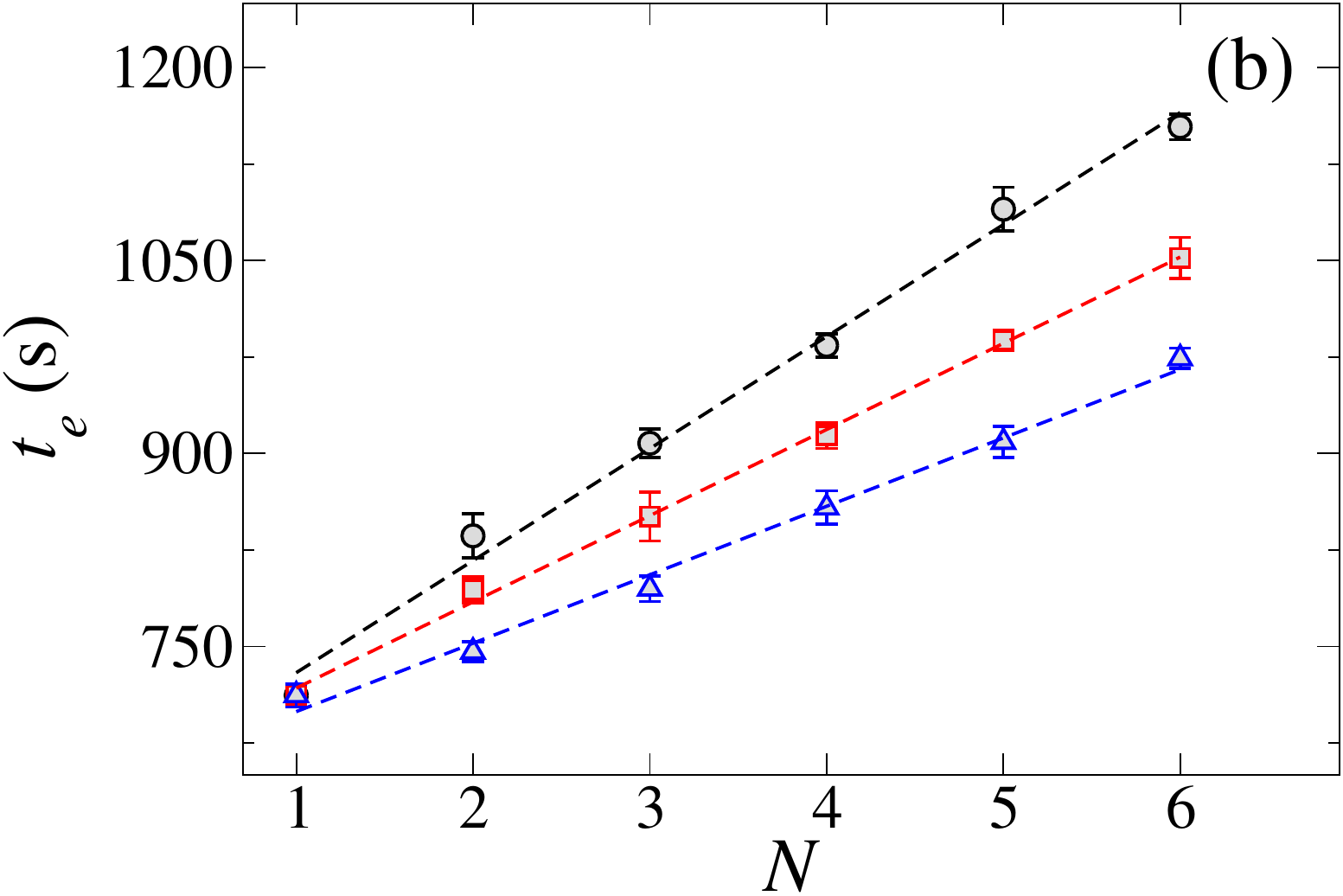} \hspace{2mm} \includegraphics[width=0.45\textwidth]{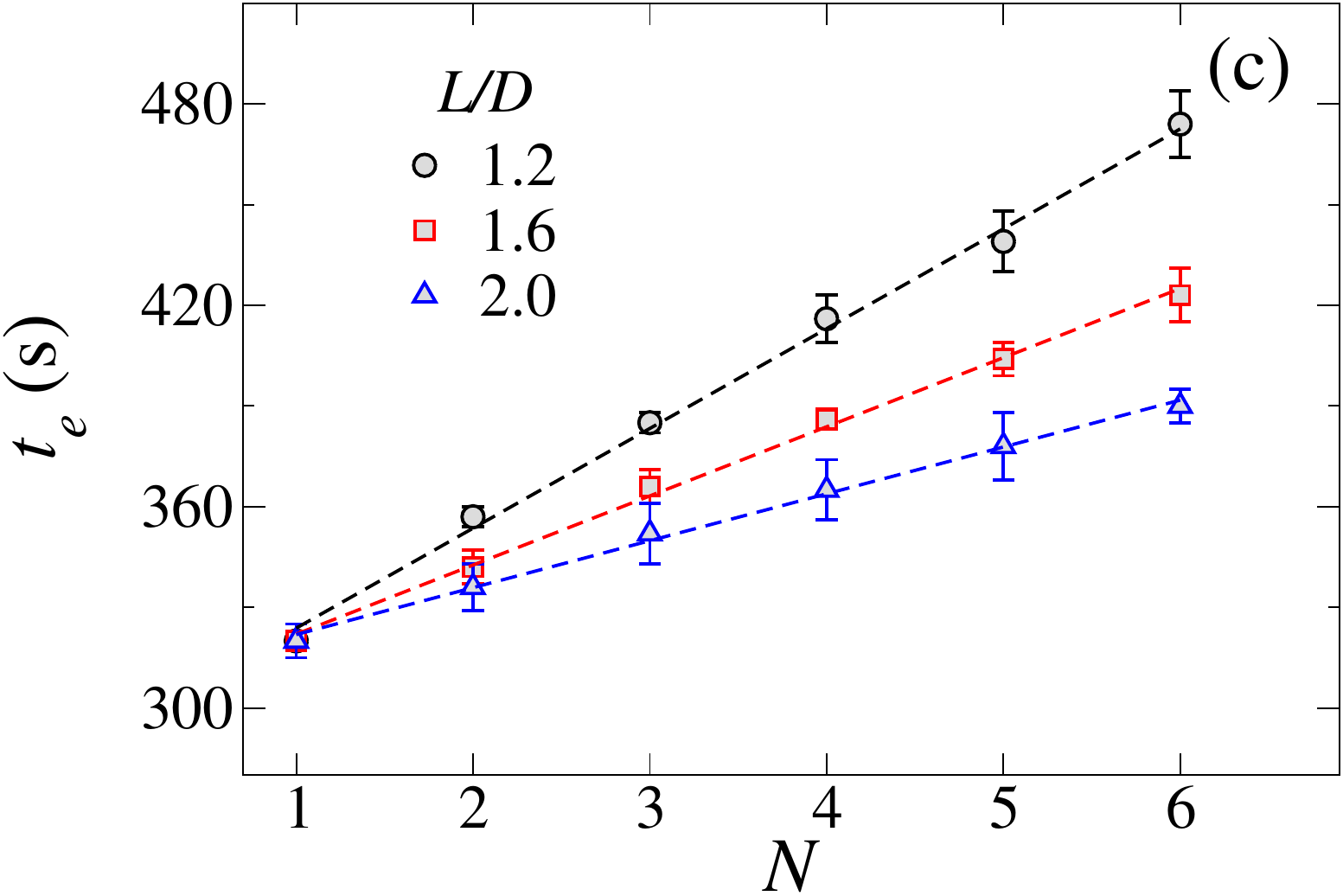}\\
\caption{Variation of the lifetime $t_e$ (in second) of the droplets with $N$ (different configurations as depicted in Fig. \ref{fig:fig2}d) for (a) $T_s = 25^\circ$C, (b) $T_s = 45^\circ$C and (c) $T_s = 65^\circ$C.}
\label{fig:fig3}
\end{figure}

The variations in lifetimes with increasing numbers of droplets are depicted in Figure \ref{fig:fig3}(a-c) for substrate temperatures of $T_s = 25^\circ$C, $45^\circ$C, and $65^\circ$C, respectively. Overall, it is evident that at all substrate temperatures, the lifetime increases with the number of droplets in the configuration. Additionally, at a given droplet configuration and temperature, the lifetime decreases with an increase in the $L/d$ ratio. However, the extent of variation in lifetimes differs across scenarios.
At a substrate temperature of $25^\circ$C, the lifetime of an isolated droplet is approximately $3720 \pm 180$ seconds (s), with a maximum lifetime increase of 91\% observed for six droplet configurations at $L/d = 1.2$. This increase is attributed to the diffusion-based interaction of vapour fields from all droplets in the configuration. As the number of droplets increases, the local vapour concentration near the evaporating droplet rises, thereby reducing the evaporation rate. This phenomenon, termed the vapour shielding effect, is documented in previous studies \cite{chen2022predicting,schofield2020shielding}.
For $L/d = 1.6$ and $L/d = 2.0$, the droplets exhibit an increase of 74\% and 57\% lifetime, respectively. As the separation distance between droplets increases, the shielding effect diminishes, resulting in slightly higher evaporation rates compared to $L/d = 1.2$. The percentage increase in the lifetimes of the central droplet compared to that of an isolated droplet for all cases is provided in Table S3 (supplementary material).

For the case of higher substrate temperatures, a maximum lifetime increase of 62\% and 48\% for six droplet configurations with $L/d = 1.2$ were obtained at  $T_s=45^\circ$C and $65^\circ$C, respectively. This is clearly lower than the increase observed for the $T_s=25^\circ$C case. The difference is due to the natural convective effects, where the heated air in contact with the substrate rises due to buoyancy, thereby altering the local vapour concentration. Thus, the increase in local vapour concentration is not as high as in low-temperature cases where diffusive effects dominate. This results in the diminishing of the shielding effect, which tends to increase the evaporation rate. This effect can also be observed from the slope values shown in Table \ref{table:T2} for various $L/d$ ratios at different substrate temperatures. These slope values are obtained by imposing a linear fit on the lifetime data of the droplets after normalizing them with the isolated droplet lifetime at the corresponding substrate temperature. The lower slope of the fitted line for the variation of the lifetime $(t_e)$ as a function of $N$ in different configurations represents a weaker effect of neighbouring droplets on the evaporation rate of the central droplet.

The increase in the lifetime of the side droplets within a specific array configuration is compared with that of an isolated droplet in Table S2. It can be seen that the lifetime of the side droplets increases due to the shielding effect. Additionally, the side droplet lifetime increases with an increase in the number of droplets in an array. Conversely, within a given configuration, the lifetime decreases with an increase in the $L/d$ ratio and substrate temperature. Essentially, the side droplets also exhibit variations in lifetime trends similar to those of the central droplet. However, the relative variations are not as pronounced as those experienced by the central droplets.

The ratio of the lifetime of the side to the central droplet in an array configuration at different temperatures is presented in Table S1 (supplementary material). It is observed that for a given number of droplets and temperature, the ratio of the lifetimes of the side to the central droplets slightly increases as the $L/d$ ratio is increased. This increase is expected as the side and central droplet lifetimes converge with higher $L/d$ ratios. Interestingly, for a fixed number of droplets and $L/d$ ratio, the ratio of side-to-central droplet lifetimes remains consistent across all substrate temperatures. However, Table S2 and Table S3 show that at a fixed number of droplets and $L/d$ ratio, the increase in lifetime relative to an isolated droplet is greater at a substrate temperature of $25^\circ$C compared to $65^\circ$C. This indicates that the shielding effect provided by the droplets is reduced at higher temperatures, affecting both the side and central droplets similarly. Therefore, the side-to-central droplet lifetime ratio is maintained consistently across all temperatures due to this scaled reduction in the shielding effect.

\begin{table}
\centering
\caption{Slopes of the best-linear fit (with $R^2 = 0.99$)  in Fig. \ref{fig:fig3} for different $L/d$ ratios and substrate temperatures ($T_s$).} 
\label{table:T2}
\hspace{2 mm}\\
\begin{tabular}{|c|l|l|}
\hline
\multicolumn{1}{|l|}{Substrate temperature $(T_s)$} & $L/d$ & Slope \\ \hline
\multirow{3}{*}{$25^\circ$C}      & $1.2$ & $0.18$          \\ \cline{2-3} 
                                  & $1.6$ & $0.14$          \\ \cline{2-3} 
                                  & $2.0$ & $0.11$          \\ \hline
\multirow{3}{*}{$45^\circ$C}      & $1.2$ & $0.12$          \\ \cline{2-3} 
                                  & $1.6$ & $0.09$          \\ \cline{2-3} 
                                  & $2.0$ & $0.07$          \\ \hline
\multirow{3}{*}{$65^\circ$C}      & $1.2$ & $0.09$          \\ \cline{2-3} 
                                  & $1.6$ & $0.06$          \\ \cline{2-3} 
                                  & $2.0$ & $0.04$          \\ \hline
\end{tabular}
\end{table}

\begin{figure}[h]
\centering
\hspace{0.5cm}\\
\hspace{-0.2cm}\includegraphics[width=1\textwidth]{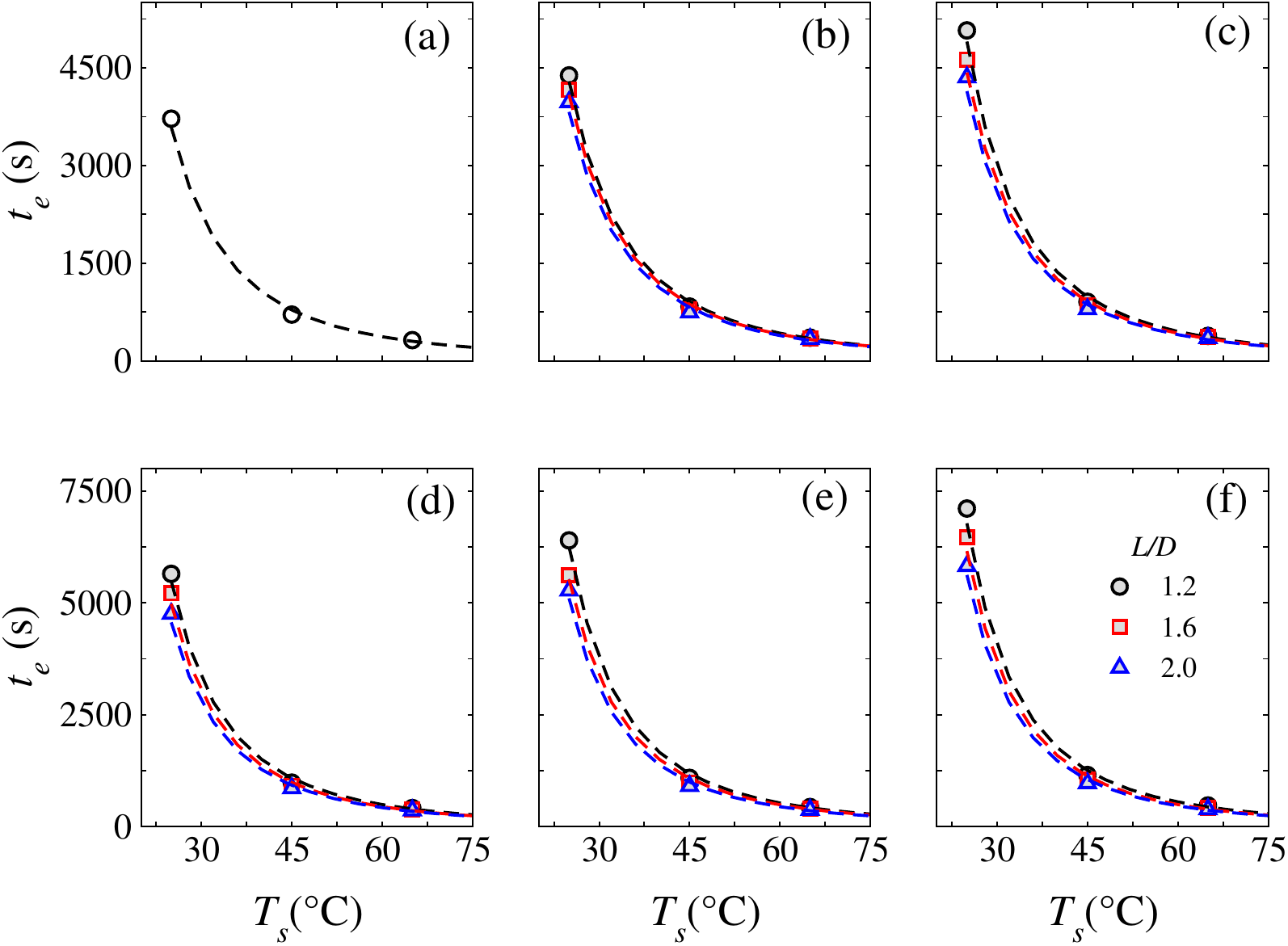} \hspace{2mm}\\
\caption{Variation of the lifetime $(t_e)$ of the droplets (in second) with substrate temperature ($T_s$) for different $L/D$ ratios and different numbers of droplets ($N$) arranged in various configurations as shown in Fig. \ref{fig:fig2}d.  (a) $N = 1$, (b) $N = 2$, (c) $N = 3$, (d) $N = 4$, (e) $N = 5$ and (f) $N = 6$.}
\label{fig:fig4}
\end{figure}

The variation of the lifetime $(t_e)$ of the droplets with substrate temperature ($T_s$) for different configurations at all $L/d$ ratios is discussed below. As substrate temperature increases, the evaporation time decreases for all configurations and $L/d$ ratios. A power-law relationship between the evaporation time and substrate temperature is observed in Figure \ref{fig:fig4}. A similar relation was proposed by \citet{girard2008influence}, which was also verified by \citet{dash2014droplet} for an isolated sessile droplet. We found that this relation holds well even for multiple droplet arrangements at different $L/d$ ratios. A power law fit of the form $t_e = aT_s^{n}$, where $t_e$ is the total evaporation time, $T_s$ is the substrate temperature, and $a$ and $n$ are fitting constants, is matched with the experimental data with reasonable accuracy. This is illustrated in Fig. \ref{fig:fig4} for various droplet configurations. The power law expressions are observed to be different for various configurations while exhibiting only slight variations across $L/d$ ratios within a given configuration.

Table S4 (supplementary material) shows that the scale factor values $a$ increase with a decrease in $L/d$ ratios for a given configuration, and at a given $L/d$ ratio, they increase with the number of droplets in the configuration. This suggests that increasing the number of droplets and decreasing the $L/d$ ratio lead to increased evaporation time, which is reflected in the scale factor value $a$. The power law fitting for the lifetime can be qualitatively understood by considering the variation of the saturation vapour pressure of water with increasing substrate temperatures. Since the evaporation rate is proportional to the difference between the vapour concentration at the liquid-air interface and the ambient, the trends in evaporation lifetime can be visualized by considering the variation of the inverse of the saturation vapour pressure difference of water between the droplet temperature and ambient temperature. Additionally, in Supplementary Figure S1, it is observed that the variation of the inverse of the saturation vapor pressure difference, when fitted with a power law, has a similar exponent to the power law fit for the lifetime variation of an isolated droplet case ($-2.53$ and $-2.59$, respectively). This similarity indicates that the lifetimes of droplets across all configurations are influenced in a similar manner. However, as the number of droplets increases, the value of the exponent rises compared to the isolated droplet case, indicating that the shielding effect has a secondary impact, perturbing the temperature dependence of droplet lifetimes.

\subsection{Evaporation dynamics: Top view}

\begin{figure}
\centering
\vspace{-1cm}\includegraphics[width=0.8\textwidth]{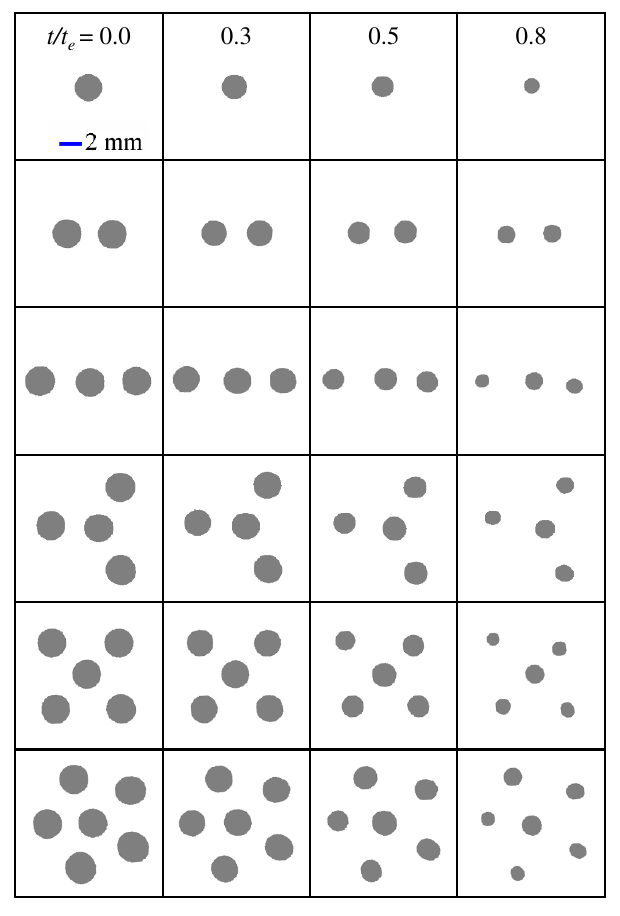} \hspace{2mm}\\
\caption{Temporal evolution of the top view profiles of the droplets with $L/d = 1.6$ at $T_s =45^\circ$C.}
\label{fig:fig5}
\end{figure}

In this section, we examine the top view of the droplets arranged in different configurations. The temporal evolution of the top view of the droplets arranged in various configurations at different normalized evaporation times for $L/d = 1.6$ at $T_s = 45^\circ$C is illustrated in Figure \ref{fig:fig5}. It can be seen that the diameter of the drops decreases continuously as time progresses, whereas the side view profiles show pinned behaviour of the droplets for a certain time after deposition (till $t/t_e = 0.3$). This is because the top view captures the equatorial diameter of the drops, which is slightly higher than the wetting diameter for droplets on hydrophobic surfaces. As the droplet evaporates, the contact angle decreases, and the equatorial diameter becomes equal to the wetting diameter for contact angles less than $90^\circ$. Even though all droplets have more or less the same initial volume, it can seen that the central droplet in all the configurations is slightly larger compared to the side droplets (except the 2-droplet configuration) for most of the evaporation time. This is due to the shielding effect of the side droplets on the central drop. The central droplet is surrounded in all directions by the side droplets, which leads to a higher vapour concentration along the azimuthal direction near the vicinity of the central drop. For the side droplets, the vapour concentration near the surface facing the central drop is higher than its opposite surface. Hence, the total vapour concentration in the azimuthal direction around these side droplets is lower and slightly non-uniform. This slight difference in the vapour concentration increases the side drop's evaporation rate, making it evaporate faster than the central droplet. The top view arrangements of the various droplet configurations at $t=0$ for different $L/d$ ratios are shown in Supplementary Figure S2.

\subsection{Evaporation dynamics: Side view}

\begin{figure}[h]
\centering
\hspace{0.1 cm}\\
\hspace{-0.2cm}\includegraphics[width=1.0\textwidth]{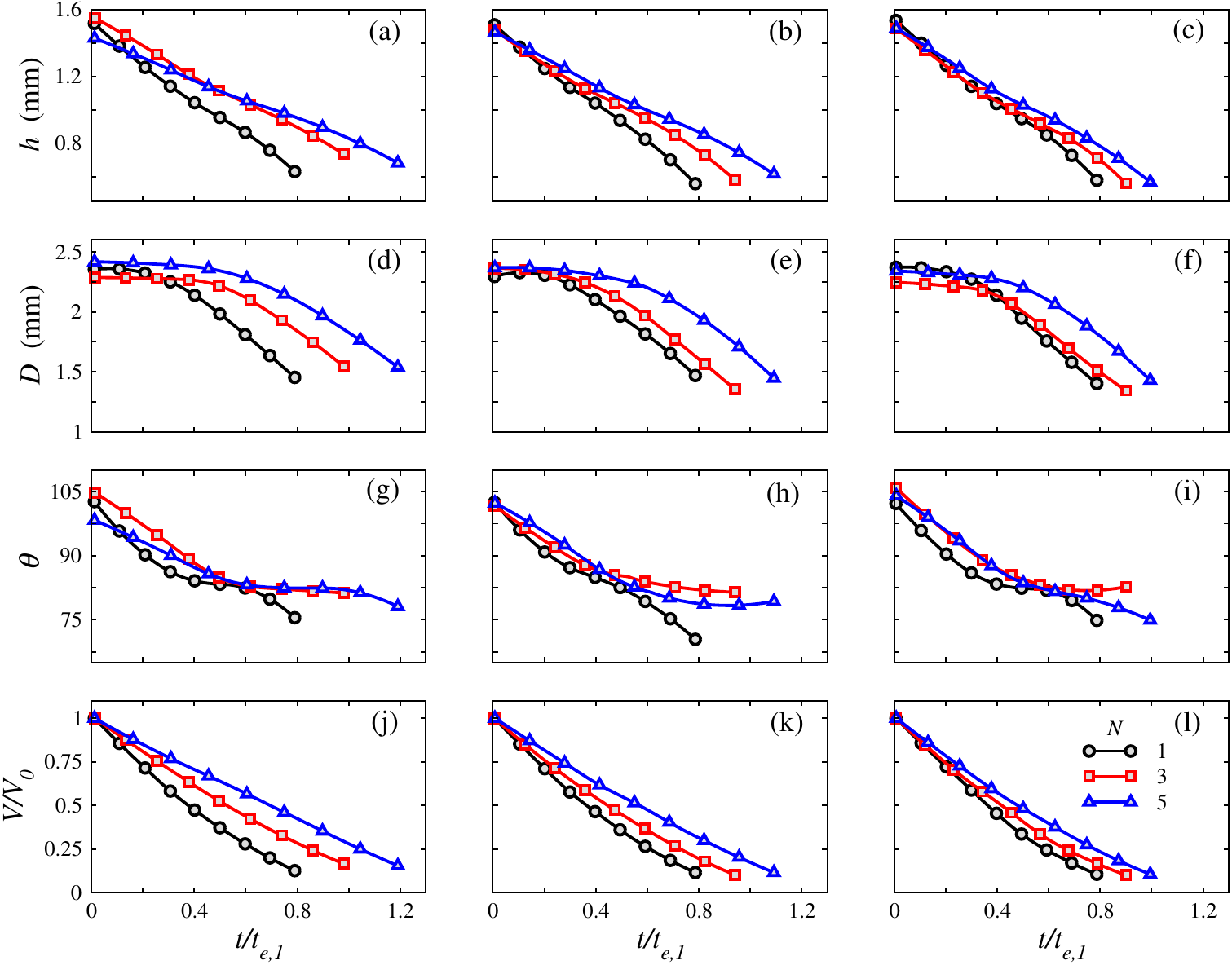} \hspace{2mm}\\
\caption{Variations of the height ($h$ in mm), wetted diameter ($D$ in mm), contact angle ($\theta$ in degree), and normalized volume ($V/V_0$) of multiple droplets with $L/d = 1.6$. The first (a, d, g, j), second (b, e, h, k), and third columns (c, f, i, l) represent the droplet parameters at a substrate temperature $T_s$ of $25^\circ$C, $45^\circ$C and $65^\circ$C, respectively.}
\label{fig:fig6}
\end{figure}

In this section, we present the temporal evolution of the geometrical parameters of the droplets, namely height ($h$ in mm), wetted diameter ($D$ in mm), contact angle ($\theta$ in degrees), and droplet volume normalized by initial volume ($V/V_0$) of the sessile water droplet. The experimental section has already discussed the details of extracting contour profiles from CMOS camera data. Since accurate estimation of droplet parameters can be obtained only up to $t/t_e = 0.8$ due to imaging limitations, the variation of these parameters is presented for the isolated droplet and the central droplet in the 3-droplet and 5-droplet configurations. Figure \ref{fig:fig6} shows the variation of the droplet parameters with the evaporation time normalized with the lifetime of an isolated droplet ($t_{e,1}$) for $L/d = 1.6$. Fig. \ref{fig:fig6}(a,d,g,j), Fig. \ref{fig:fig6}(b,e,h,k), and Fig. \ref{fig:fig6}(c,f,i,l) show the temporal variations of the droplet geometrical parameters at a substrate temperature of $25^\circ$C, $45^\circ$C, and $65^\circ$C, respectively. For all substrate temperatures, the height $(h)$ and normalized volume $(V/V_0)$ of the droplet decrease monotonically with time. The variation of the wetted diameter ($D$) reveals that the droplet evaporates in two modes, namely in constant contact radius (CCR) during the initial phase of its lifetime, followed by a monotonic decrease, in which constant contact angle (CCA) mode is prevalent. This CCA mode is inferred from the near constancy in contact angle in the intermediate evaporation stages. After this, near the end stage, a mixed mode of evaporation is sometimes observed, where both the wetted diameter and contact angle decrease. As the number of droplets in the configurations increases, the time associated with the CCR and CCA mode also increases, as seen from Fig. \ref{fig:fig6}(d-g). The droplet in the 5-droplet configuration is pinned for a longer time compared to that of an isolated droplet. Consequently, the change in the droplet parameters with time is slower for configurations with more droplets.

A similar trend is observed for droplets at higher substrate temperatures. Although the relative increase in the duration of evaporation modes (CCR and CCA) remains the same at all temperatures, the actual duration increase with respect to that of an isolated droplet at the corresponding temperature is reduced (the difference in the CCR duration between the 5-droplet and isolated droplet configuration at $25^\circ$C, $45^\circ$C, and $65^\circ$C is around 370, 71 and 32 seconds respectively). This can be seen from the fact that at high temperatures, the curves of 1, 3, and 5 droplet configurations come closer to each other across all droplet parameters. At $65^\circ$C, the curves are much closer when compared with those of $45^\circ$C, indicating the diminishing effects of neighbouring droplets on the droplet parameters. Thus, the difference in the variation of the height, wetted diameter, and normalized volume among the droplet configurations decreases as the substrate temperature increases.  

At a given temperature, as the $L/d$ ratio is increased to $2.0$, the variations in the parameters among the configurations decrease and tend to approach that of the isolated droplet trend. The lifetime of a 5-droplet configuration at $25^\circ$C is around $1.6$ times the isolated droplet with $L/d = 1.6$, whereas it is reduced to $1.5$ times with an $L/d = 2.0$. A similar trend is observed for $L/d = 2.0$ at high temperatures; the lifetime of the central droplet in the 5-droplet configuration is just $1.2$ times the isolated droplet at $65^\circ$C. The variations of the droplet geometrical parameters $(h,D,\theta,V/V_0)$ with $L/d = 2.0$ are shown in Supplementary Figure S3. Extending this argument, we can say that, at a particular temperature and configuration, there is a value of $L/d$ ratio for which the lifetime of the multiple droplets becomes equal to the lifetime of an isolated droplet. The evaporating droplets in the group don't influence each other and behave as individual droplets. It can also be seen that this value of the $L/d$ ratio, where the drops no longer influence each other, will become lower at higher substrate temperatures. Finally, we could say that all the changes observed boil down to the alterations in the local vapour concentration around the droplets (shielding effect), which in turn is controlled by the number of droplets in configuration, $L/d$ ratios, and the substrate temperature.

\subsection{Generalised behaviour}

\begin{figure}[h]
\centering
\hspace{0.1cm}\\
\hspace{-0.2cm}\includegraphics[width=0.9\textwidth]{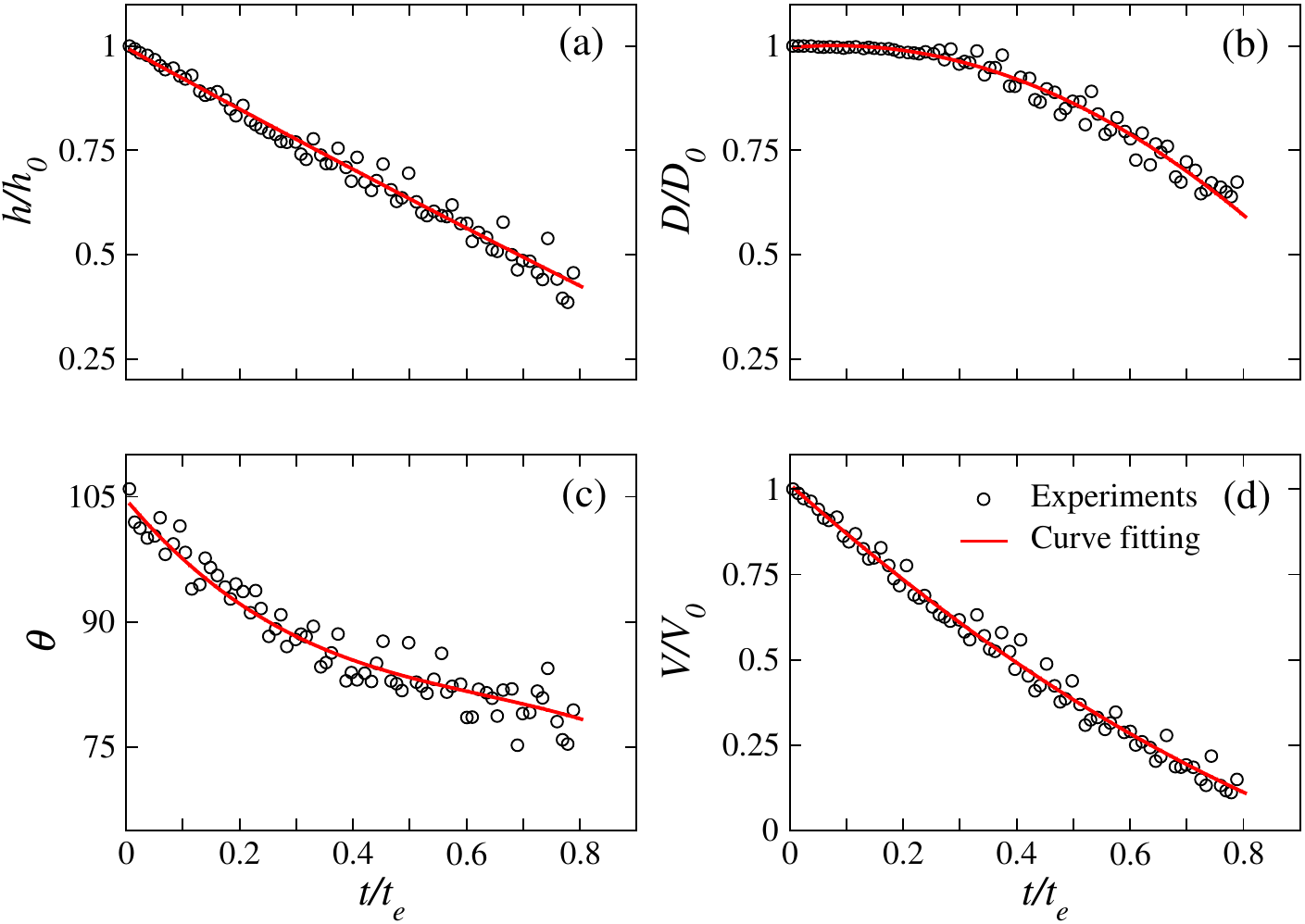} \hspace{2mm}\\
\caption{Variations of (a) normalized height ($h/h_0$), (b) normalized wetted diameter ($D/D_0$), (c) contact angle ($\theta$ in degree), and (d) normalized volume ($V/V_0$) of multiple droplets with normalized evaporation time $t/t_e$. The average values of $h_0$, $D_0$, $\theta_0$, and $V_0$ are $1.48$  mm, $2.32$  mm, $104.62^\circ$, and $4.83$ mm$^3$ respectively. The results for 1-, 3-, and 5-droplet configurations with $L/d$ ratios of 1.6 and 2.0 at the three different temperatures are included.}
\label{fig:fig7}
\end{figure}

In the previous section, the variations of the central droplet parameters for different configurations are obtained by normalizing the time with the corresponding isolated droplet lifetime ($t_{e,1}$). This approach highlights the differences associated with droplet configuration, $L/d$ ratio, and substrate temperatures, as seen in Fig. \ref{fig:fig6}. We found a generalized behaviour when plotting these results on normalized scales. Figures \ref{fig:fig7}(a-d) depict the variations of normalized height $(h/h_0)$, wetted diameter $(D/D_0)$, contact angle $(\theta)$, and normalized volume with $t/t_e$, where $h_0$ and $D_0$ represent the initial height and wetted diameter of the droplet, respectively. Fig. \ref{fig:fig7} presents the experimental data points (black circles) of a single isolated, 3 and 5 droplet configurations with $L/d$ ratios of $1.6$ and $2.0$ at different substrate temperatures of $25^\circ$C, $45^\circ$C and $65^\circ$C normalized with the lifetime of each configuration taken from the Table \ref{table:T1}. A curve fitting was done on the experimental data points to get the average behaviour (red line) across all the droplet parameters. The parameters associated with the polynomial fit for $h$, $D$, $\theta$ and $V$ are provided in Supplementary Table S5, along with the $R^2$ values, which denote the deviation of the experimental data points from the average value. It could be seen that most of the experimental values lie close to the average value. Hence, we can say that in all cases with different droplet configurations, $L/d$ ratios, and temperatures, the contact line dynamics behave similarly when normalizing with their lifetimes. \citet{dash2014droplet} also found this behaviour in the evaporation of an isolated sessile water droplet on hydrophobic and superhydrophobic surfaces at high temperatures. \citet{hatte2019universal} also reported a similar behaviour for 3-droplet configuration with $L/d$ ratios ranging from 1.1 to 2.6 for droplets evaporating at room temperature conditions. We extend this behaviour by incorporating multiple droplet arrays with $L/d$ ratios of $1.2$, $1.6$, and $2.0$ for high substrate temperatures. The fact that similar profiles were obtained for various droplet configurations and substrate temperatures shows that the impact of neighbouring droplets or substrate temperature is limited to changes in the evaporation rate and does not affect the droplet shape evolution dynamics for a given substrate and fluid combination. Hence, only the total evaporation time is altered across the various conditions investigated here, and a generalized behaviour can be obtained by normalizing with the droplet lifetime. It could be said that the changes in local vapour concentration brought about by either changing the number of droplets in the configuration, $L/d$ ratios, or substrate temperature do not significantly alter the contact line dynamics. However, changes in the internal flow dynamics were observed by examining the deposition drying pattern\cite{pradhan2015deposition}.

\subsection{Theoretical modeling}
This section compares our experimental results with the predictions made using the theoretical model reported in Refs.  \cite{wray2020competitive,masoud2021evaporation}. As the diffusion-based model overestimates the evaporation rates of water droplets on a hydrophobic surface under high substrate temperatures \cite{sobac2012thermal}, we have integrated the evaporative cooling mechanism into the theoretical model proposed by \citet{wray2020competitive}. This modified theoretical model now enables us to predict the instantaneous evaporation rates of the droplets more accurately. For all our cases, the contact radius of the droplet is less than the capillary length scale, so the effect of gravity on the droplet shape can be neglected. The typical value of the Bond number $(\rho g R^2/ \sigma)$ for the droplets considered in our study is around 0.3. Thus, a spherical cap profile is used to calculate the volume during the evaporation process. Further, it is assumed that the vapor concentration at the liquid-air interface is at the saturated condition  $C_{sat}\left ( T_{drop} \right)$, where $T_{drop}$ is the surface averaged droplet temperature after taking evaporative cooling into account. The vapor concentration in the ambient far from the droplet is  $HC_{sat}\left ( T_{\infty} \right )$, where $H$ and $T_{\infty}$ are the ambient relative humidity and temperature, respectively. Using infrared imaging techniques, we calculate the evaporative cooling effect of the droplet, which results in a reduction of the droplet temperature. Initially, we obtain the temperature profile of an isolated evaporating droplet at a given substrate temperature. Subsequently, this temperature profile is averaged over the droplet's surface area to determine a uniform reduced surface temperature of the droplet ($T_{drop}$). Supplementary Figure S4 displays the IR images of isolated droplets and their corresponding temperature variations across radial locations at different substrate temperatures. When considering only the diffusion model and assuming the droplet to be at the substrate temperature, the evaporation rates for an isolated droplet were found to be overestimated by approximately $4\%$, $18\%$, and $39\%$ at substrate temperatures of $25^\circ$C, $45^\circ$C, and $65^\circ$C, respectively. However, upon incorporating evaporative cooling effects for an isolated droplet, these deviations reduced to $0.6\%$, $2.8\%$, and $6.2\%$ at the corresponding substrate temperatures. In the case of droplets within array configurations, thermal imaging reveals slightly elevated temperature profiles due to reflections between the droplets. Consequently, the calculated evaporation rates assume that the temperature profiles for multiple droplets are similar to that of an isolated droplet. The instantaneous evaporation rate of an isolated droplet is given by \cite{hu2014effect}

\begin{equation}\label{j:eq1}
\dot{m}_{iso} = \pi RD_{v}M\left [ C_{sat}\left ( T_{drop} \right )- HC_{sat}\left ( T_{\infty} \right )\right]f\left ( \theta  \right ),
\end{equation}
where
\begin{equation}
f \left ( \theta  \right ) = \frac{\sin \theta }{1+\cos \theta } + 4\int_{0}^{\infty}\frac{1+\cosh 2\theta \tau }{\sinh 2\pi \tau }\tanh \left [ \left ( \pi -\theta  \right ) \tau \right ]d\tau,
 \end{equation}
$R$ is the contact radius of the droplet and $D_{v}$ is the vapor diffusion coefficient at the mean temperature $(T_{s} + T_{\infty})/2$. Moreover, for any droplets arranged in an array configuration with arbitrary geometry and spacing, the evaporation rate of the droplets in the array with $N$ droplets can be determined as \cite{wray2020competitive}
\begin{equation} \label{j:eq2}
\dot{m}_{o} = \dot{m}_{iso} - \frac{2}{\pi }\sum_{k=1}^{N}\dot{m}_{s,k}\arcsin\left (\frac{R_{o}}{D_{s,k}}\right),
\end{equation}
where, $\dot{m}_{o}$, $\dot{m}_{iso}$, and $\dot{m}_{s,k}$ denote the evaporation rates of the reference, isolated, and surrounding droplets, respectively; $R_{o}$ represents the contact radius of the reference droplet and $D_{s,k}$ stands for the distance between the centers of two adjacent droplets in the array. To determine the lifetime of the droplets, Eq. (\ref{j:eq2}) is solved concurrently for a system of $N \times N$ linear equations. Here, Eq. (\ref{j:eq2}) can be rearranged into a matrix form as \cite{edwards2021interferometric}.
\begin{equation} \label{j:eq3}
\dot{m}_{o} = \phi ^{-1}\dot{m}_{iso},
\end{equation}
where $\phi ^{-1}$ is known as the suppression matrix of the order $N \times N$. The off-diagonal elements in the suppression matrix capture the interaction between the droplets in the array, while the diagonal elements capture the dynamics of the individual droplet. For the configurations examined in this study, when $N=2$, Eq. (\ref{j:eq2}) simplifies to a single equation that directly provides the reduced evaporation rate. However, For $N \geq 3$, due to the symmetrical arrangement of configurations (one central droplet symmetrically surrounded by side droplets), Eq. (\ref{j:eq2}) can be reduced to two equations that are solved simultaneously. The suppression matrix is of size $2\times2$ for these cases. Using this approach, we theoretically predicted the evaporation rates of the central and side droplets and their lifetimes. It is to be noted that we can estimate the contact radius and contact angle of the central droplet accurately up to $t/t_e = 0.8$ due to their resultant complex morphologies at $t/t_e > 0.8$. Thus, we employ the theoretical model to estimate the evaporation time up to 80\% of the total lifetime (denoted by $t_{e,80}$). 

\begin{figure}
\centering
\includegraphics[width=0.9\textwidth]{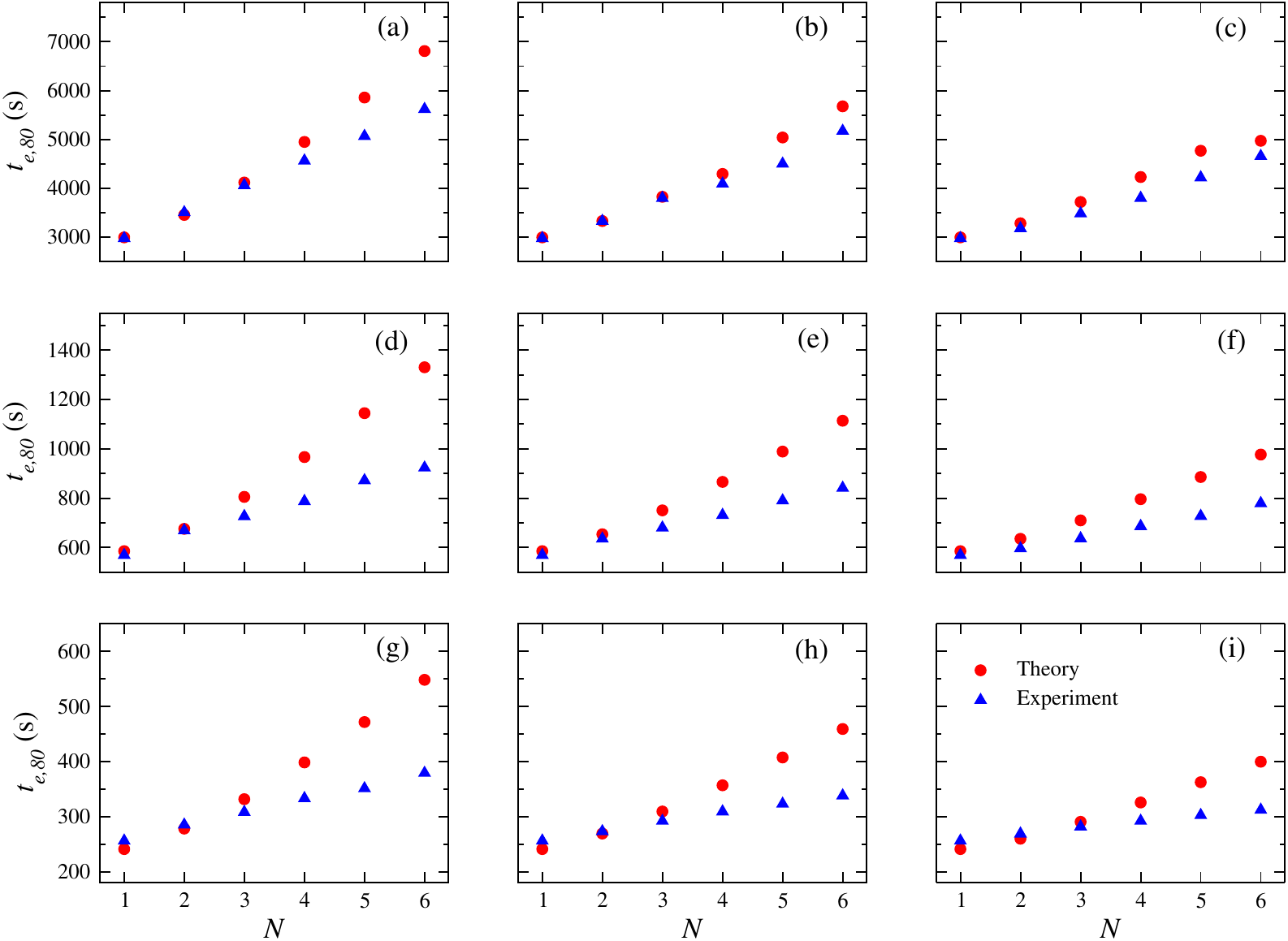}
\caption{Comparisons between the variation of the experimentally obtained 80\% lifetimes of the central droplet $(t_{e,80})$ and theoretical predictions with versus number of droplets ($N$) for different substrate temperatures. Panels (a, b, c), (d, e, f) and (g, h, i) correspond to $T_s=25^\circ$C, $T_s=45^\circ$C and $T_s=65^\circ$C, respectively. The panels (a, d, g), (b ,e, h) and (c, f, i) are associated with $L/d=1.2$, $L/d=1.6$ and $L/d=2.0$ respectively.}
\label{fig:fig8}
\end{figure}

\begin{figure}
\centering
\includegraphics[width=0.9\textwidth]{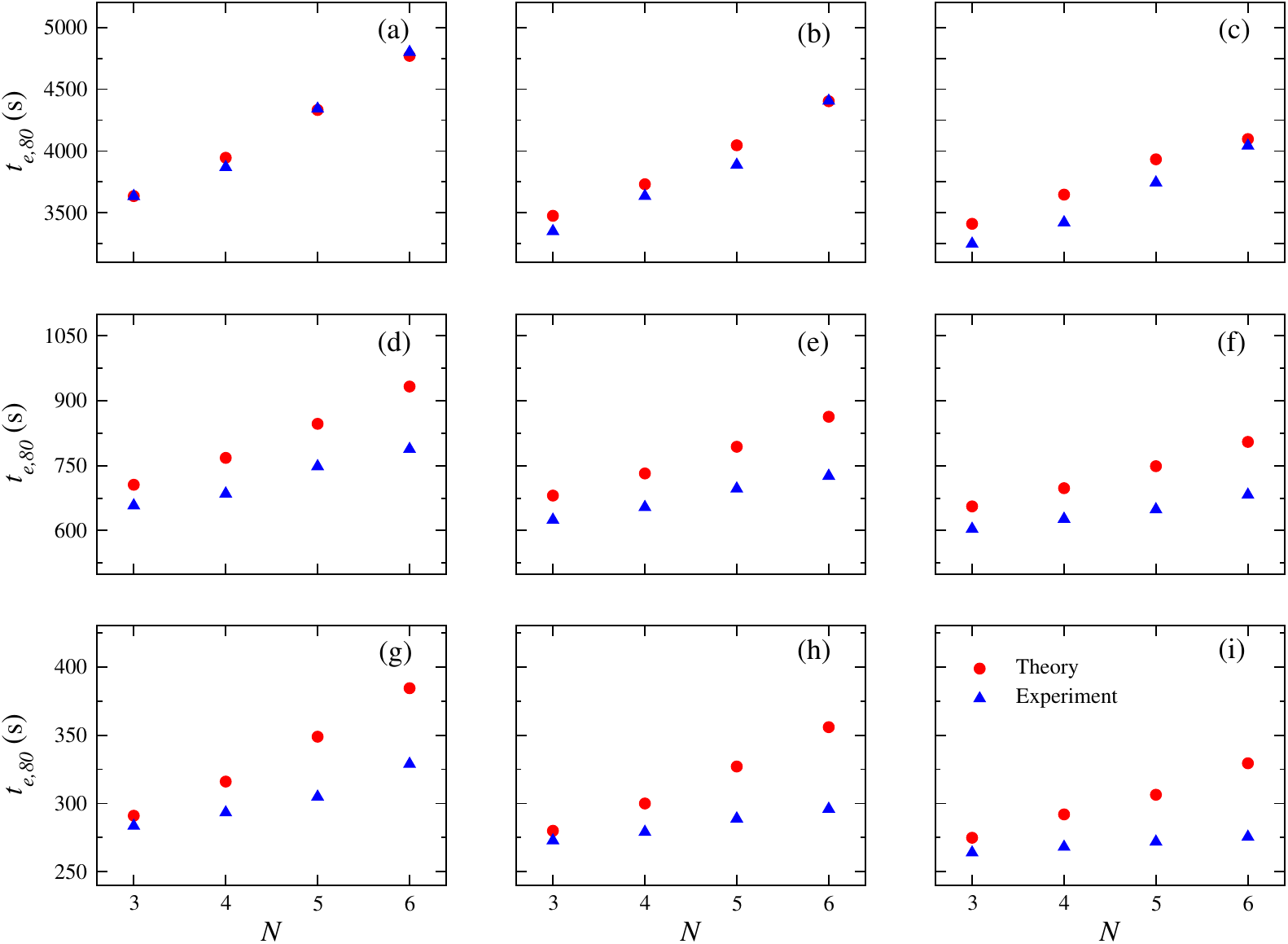}
\caption{Comparisons between the variation of the experimentally obtained 80\% lifetimes of the side droplets $(t_{e,80})$ and theoretical predictions with versus number of droplets ($N$) for different substrate temperatures. Panels (a, b, c), (d, e, f) and (g, h, i) correspond to $T_s=25^\circ$C, $T_s=45^\circ$C and $T_s=65^\circ$C, respectively. The panels (a, d, g), (b ,e, h) and (c, f, i) are associated with $L/d=1.2$, $L/d=1.6$ and $L/d=2.0$ respectively.}
\label{fig:fig9}
\end{figure}

Figures \ref{fig:fig8} and \ref{fig:fig9} depict the comparisons between the experimentally obtained $t_{e,80}$ and theoretical predictions of central and side droplet lifetimes, respectively, for different substrate temperatures and $L/D$ values in various configurations. The theoretical model predicts the experimental results quite satisfactorily for $T_s=25^\circ$C for the side droplets, whereas, for the central droplets, we observe slight variations at higher numbers of droplets. The deviation between the theoretical prediction and experimental result is given by:
\begin{equation} \label{j:eq4}
Error = \frac{t_{e,80_{theory}} - t_{e,80_{experiment}}}{t_{e,80_{theory}}}.
\end{equation}

It is observed that for the side droplets, the maximum error is about 6.2\% for the four-droplet configuration with $L/d = 2.0$ at a substrate temperature of $25^\circ$C. For all other scenarios, the error remained below 5\%. For the central droplets, the errors are 13.5\% and 17.5\% for five- and six-droplet configurations with $L/d = 1.2$. For all other scenarios, the error remained below 10\%. The deviation between the theoretical predictions and experimental observations may result from implementing the model for high-contact angle experiments, as this model is strictly valid for low-contact angles. \citet{wray2020competitive} also employed this model and compared it with the evaporation rates of \citet{khilifi2019study}, conducted with contact angles around $80^\circ$, and obtained good agreement for a seven-droplet configuration. Our study shows a significant disparity between experimental and theoretically predicted lifetime values at higher substrate temperatures. For instance, the predicted values exhibit a discrepancy of nearly 30\% and 16\% for the central and side droplets, respectively, in the case of six-droplet configurations with $L/d = 1.2$ at substrate temperatures of $45^\circ$C and $65^\circ$C. This observed deviation is attributed to convective effects, which alter the vapor concentration gradients around the droplets within the array.

The influence of convective effects is commonly characterised using a dimensionless parameter known as the Rayleigh number, denoted as $Ra$, which is expressed as:
\begin{equation} \label{j:eq5}
Ra = \frac{g\beta \Delta TR_{c}^3}{\gamma \alpha},
\end{equation}
Here, $g$ represents the acceleration due to gravity, $\beta$ is the thermal expansion coefficient, $\Delta T$ signifies the temperature difference between the substrate and the surrounding environment, $R_{c}$ denotes the characteristic length, while $\gamma$ and $\alpha$ stand for the kinematic viscosity and thermal diffusivity of air, respectively. The characteristic length, $R_{c}$, is determined as the contact radius for an isolated droplet. In the case of multiple droplet configurations, $R_c$ represents the radius of the array. It denotes the distance from the center of the centrally placed droplet to the outer edge of any side droplet within that array. The fluid properties of air at various substrate temperatures are listed in Supplementary Table S6. In our study, the Rayleigh numbers for isolated droplets at substrate temperatures of $25^\circ$C, $45^\circ$C, and $65^\circ$C are calculated to be 0.67, 3.75, and 6.16, respectively. In the case of a two-droplet arrangement with a ratio of characteristic length to droplet diameter ($L/d$) equal to 1.2, the corresponding Rayleigh numbers at substrate temperatures of $25^\circ$C, $45^\circ$C, and $65^\circ$C are 7.1, 40, and 65.5, respectively. For configurations involving three or more droplets with $L/d = 1.2$, the Rayleigh numbers at substrate temperatures of $25^\circ$C, $45^\circ$C, and $65^\circ$C are found to be 26, 147, and 242, respectively. At substrate temperatures of $45^\circ$C and $65^\circ$C, the Rayleigh numbers are $5.6$ and $9.2$ times higher compared to the $25^\circ$C case. Consequently, higher temperatures lead to increased upward air motion. This enhanced motion results in the easier removal of water vapor and air from the vicinity of the droplets. Thus, there is reduced interaction between the droplets and vapor within the vapor confinement region (the space between the droplets in an array), impacting the shielding effect experienced by the droplets. Due to this reduced shielding, the increase in vapor concentration is not as pronounced as observed at lower substrate temperatures. The values of the Rayleigh numbers for droplets with various configurations and $L/d$ ratios at different substrate temperatures are provided in Supplementary Table S7. The numerical investigation by \citet{ait2010numerical} illustrates vapor concentration distribution velocity fields around an isolated droplet at both room and higher substrate temperatures. They observed that while iso-concentration curves around a droplet at room temperature follow a radial pattern, this distribution differs under high substrate temperature conditions. They also observed an increase in upward air motion at a higher substrate temperature ($70^\circ$C with $Ra = 84$). Consequently, droplets at room temperature exhibit stronger interactions with others in the array compared to those at higher substrate temperatures. At elevated temperatures, the diffusive model cannot accurately capture the interaction among droplets, which is influenced by convective effects altering vapor concentration. Thus, a comprehensive understanding of the distribution of iso-concentration curves around the droplets is imperative for accurately predicting the lifetime of multiple droplets at high substrate temperatures.

\section{Conclusions}\label{sec:conclusions}

We have performed experimental investigations to understand the evaporation dynamics of an array of sessile droplets with six different configurations, characterized by $L/d$ ratios of $1.2$, $1.6$, and $2.0$ and three distinct substrate temperatures of $25^\circ$C, $45^\circ$C, and $65^\circ$C. Utilizing shadowgraphy and infrared imaging techniques, coupled with the in-house post-processing method employed in \textsc{Matlab}$^{\circledR}$, we obtained side and top view profiles to track the evolution of height, spread, contact angle, and volume. The lifetime of an array of droplets is observed to surpass that of an isolated droplet due to the shielding effect induced by neighbouring droplets, which elevates the local vapour concentration, consequently reducing the evaporation rate. At a fixed configuration and substrate temperature, lifetimes increase as droplet separation distance decreases. Moreover, lifetimes increase with the number of droplets in the group. Additionally, a decrease in lifetimes is noted, following a power law trend with increasing substrate temperature, with the shielding effect diminishing at higher substrate temperatures due to natural convective effects. Furthermore, we observe a generalised behaviour for the centrally placed droplet in 3- and 5-droplet configurations across all $L/d$ ratios and substrate temperatures. This arises from changes in droplet configurations, $L/d$ ratios, and substrate temperatures, ultimately modifying the local vapour concentration around the droplets without significantly altering the contact line dynamics. Furthermore, a theoretical model considering both diffusive and evaporative cooling effects is employed to compare with the experimental observations. We found that the theoretical prediction of the evaporation lifetime demonstrates a good agreement with experimental lifetime values at a substrate temperature of $25^\circ$C. However, the model overpredicts the evaporation lifetime at higher substrate temperatures of $45^\circ$C and $65^\circ$C. \\ 

\noindent{\bf Credit authorship contribution statement} 

Hari Govindha performed the experiments. All the authors contributed to the analysis of the results and to the preparation of manuscript. The project was coordinated by Kirti Chandra Sahu.\\

\noindent{\bf Declaration of Competing Interest} 

The authors declare that there is no conflict of interest.\\

\noindent{\bf Supplementary material}

\begin{itemize}
\item Table S1: The ratio of the lifetimes of side-to-central droplets in various configurations for different values of $L/d$ and substrate temperature $(T_s)$.
\item Table S2: Percentage increase in the lifetime of the central droplet within an array compared to that of an isolated droplet in different configurations, considering different values of $L/d$ and substrate temperature $(T_s)$. 
\item Table S3: Percentage increase in the lifetime of a side droplet within an array compared to that of an isolated droplet in different configurations, considering different values of $L/d$ and substrate temperature $(T_s)$.
\item Table S4: The parameters of the power law fit ($t_e = aT_s^{n}$ with $R^2 = 0.99$) obtained for different configurations and $L/d$ ratios.
\item Table S5: Parameters associated with the polynomial fits for $h$, $D$, $\theta$ and $V$ in Fig. \ref{fig:fig7}.
\item Table S6: The values of the fluid properties air at various substrate temperatures.
\item Table S7: The values of the Rayleigh number $(Ra)$ calculated in various droplet configurations with different $L/d$ ratios at different substrate temperatures.
\item Figure S1: Variation of the inverse of the saturation vapor pressure difference with temperature. The fitted power law equation is given by $2652T_s^{-2.53}$.
\item Figure S2: Top view of the various droplet configurations at $t = 0$ for different $L/d$ ratios.
\item Figure S3: Variations of the height ($h$ in mm), wetted diameter ($D$ in mm), contact angle ($\theta$ in degree), and normalized volume ($V/V_0$) for different configurations with $L/d = 2.0$. The first (a, d, g, j), second (b, e, h, k), and third columns (c, f, i, l) represent the droplet parameters at $T_s=25^\circ$C, $T_s=45^\circ$C and $T_s=65^\circ$C, respectively.
\item Figure S4: Droplet temperature distribution at different substrate temperatures. Panels (a), (c), and (e) display the IR images of an isolated droplet at $T_s=25^\circ$C, $T_s=45^\circ$C, and $T_s=65^\circ$C, respectively. Panels (b), (d), and (f) depict the corresponding temperature variation along the lines marked in blue in panels (a), (c), and (e) at different values of $t/t_e$ for $T_s=25^\circ$C, $T_s=45^\circ$C, and $T_s=65^\circ$C, respectively. Here, $r/R = 0$ is the droplet apex, while $r/R = 1$ and $r/R = -1$ are the two endpoints on the triple contact line.
\end{itemize}

\noindent{\bf Acknowledgement:} {The financial support from Science \& Engineering Research Board, India, through the grant number CRG/2020/000507 is gratefully acknowledged.}

\providecommand{\latin}[1]{#1}
\makeatletter
\providecommand{\doi}
  {\begingroup\let\do\@makeother\dospecials
  \catcode`\{=1 \catcode`\}=2 \doi@aux}
\providecommand{\doi@aux}[1]{\endgroup\texttt{#1}}
\makeatother
\providecommand*\mcitethebibliography{\thebibliography}
\csname @ifundefined\endcsname{endmcitethebibliography}
  {\let\endmcitethebibliography\endthebibliography}{}

\end{document}